\documentclass[a4paper,11pt]{article}

\usepackage{amsmath,amsfonts,amssymb,caption,graphicx}
\usepackage{ushort}
\usepackage{makeidx}
\usepackage{float}
\usepackage[numbers]{natbib}
\usepackage{accents}
\usepackage{color}
\usepackage{framed}
\usepackage{versions}
\usepackage{xcolor}
\usepackage{emptypage}
\usepackage{wrapfig}
\usepackage[T1]{fontenc}
\usepackage[utf8]{inputenc}
\usepackage{titlesec}
\usepackage{fancyhdr}
\usepackage{extramarks}
\usepackage[bookmarks]{hyperref}
\usepackage{hyperref}
\usepackage{bbm}
\usepackage[baseline]{euflag}

\hypersetup{
    colorlinks=true,   	
    linkcolor=red,      
    citecolor = [rgb]{0 0.7 0},   	
    filecolor=magenta, 	
    urlcolor=blue
}

\newcommand{\RL}{{\mathbb R}}

\newcommand{\IND}{{\mathbb I}}
\newcommand{\BBP}{{\mathbb P}}

\newcommand{\BBE}{{\mathbb E}}

\newcommand{\VAR}{\mbox{\rm Var}}

\newcommand{\Bern}{\mbox{\rm Bern}}

\def\ba{\begin{align}}
\def\ea{\end{align}}
\def\ban{\begin{align*}}
\def\ean{\end{align*}}

\def\be{\begin{eqnarray}}
\def\ee{\end{eqnarray}}
\def\ben{\begin{eqnarray*}}
\def\een{\end{eqnarray*}}

\def\bqq{\begin{equation}}
\def\eqq{\end{equation}}
\def\bqqn{\begin{equation*}}
\def\eqqn{\end{equation*}}

\def\sq{$\Box$}

\def\qed{\ifmmode\sq\else{\unskip\nobreak\hfil
\penalty50\hskip1em\null\nobreak\hfil\sq
\parfillskip=0pt\finalhyphendemerits=0\endgraf}\fi\par\medbreak}

\newsavebox{\junk}
\savebox{\junk}[1.6mm]{\hbox{$|\!|\!|$}}

\def\til={{\widetilde =}}

\def\clA{{\cal A}}

 \def\eq#1/{(\ref{#1})}

\newtheorem{theorem}{Theorem}[section]

\newtheorem{proposition}[theorem]{Proposition}
\newtheorem{lemma}[theorem]{Lemma}

\def\eq#1/{(\ref{e:#1})}

\def\bdes{\begin{description}}
\def\edes{\end{description}}

\def\notes#1{}

\definecolor{mag}{rgb}{0.7,0,0.3}
\definecolor{dgreen}{rgb}{0.1,0.5,0.1}
\definecolor{dred}{rgb}{.8,0,0}
\definecolor{gray}{rgb}{.8,.8,.8}
\definecolor{brown}{rgb}{0.6451,0.3706,0.1745}

\begin{document}
 
\title{
Finite-sample expansions for the optimal error probability\\
in asymmetric binary hypothesis testing}

\author{
Valentinian Lungu 
	\thanks{Statistical Laboratory, Centre for Mathematical Sciences, 
	University of Cambridge, Wilberforce Road, Cambridge CB3 0WB, 
	UK. Email: \texttt{vml26@cam.ac.uk}. Supported by the 
	Heilbronn Institute for Mathematical Research.}
\and 
Ioannis Kontoyiannis 
	\thanks{Statistical Laboratory, Centre for 
	Mathematical Sciences, University of Cambridge, Wilberforce Road, 
	Cambridge CB3 0WB, UK. Email: \texttt{yiannis@maths.cam.ac.uk}.}
}
\date{\today}

\maketitle

\begin{abstract}
This is an expository note, not intended
for publication.\footnote{A preliminary
version of this paper appears at ISIT 
2024~\cite{lungu-K:ISIT24}.}
The problem of binary hypothesis testing
between two probability measures is considered.
Sharp bounds are derived for the 
best achievable error probability of such tests
based on independent and identically distributed 
observations. 
Specifically, the asymmetric
version of the problem is examined, where different
requirements are placed on the two error probabilities.
Accurate nonasymptotic expansions 
with explicit constants are obtained
for the error probability, using
tools from large deviations and Gaussian approximation.
Examples are shown indicating that, in the asymmetric
regime, the approximations suggested by these
bounds are significantly more accurate
than the approximations provided by either 
normal approximation or error exponents.
\end{abstract}


\section{Introduction}
\subsection{Binary hypothesis testing}

Let $P,Q$ be two probability measures on a measurable
space $(A,\clA)$, and let $\lambda$ denote a dominating
$\sigma$-finite measure so that both $P$ and $Q$ are absolutely
continuous with respect to $\lambda$. Recall that such a $\lambda$
always exists; e.g., we can take $\lambda=P+Q$.

We revisit the binary hypothesis testing problem 
between $P$ and $Q$. The problem formulation is
simple and quite elementary, though its importance
can hardly be overstated, in view of its application
across the sciences and engineering.

A deterministic test between $P$ and $Q$ consists
of a decision region $B\subset A$, such that,
given a sample $x\in A$,
the test declares $P$ to be the `true' underlying
distribution iff $x \in B$. Therefore, 
a deterministic test can always be expressed
as the {\em deterministic} binary value of
a decision function~$\delta$,
\begin{equation*}
  \delta(x) = 
    \begin{cases}
      0 & \text{if } x \in  B,\\
      1 & \text{if } x \in  B^c.\\
    \end{cases}       
\end{equation*}
More generally,
a randomised test is a family of probabilistic kernels
$P_{Z|X}(z|x)$, $z\in\{0,1\}$, $x\in A$,
where, for each $x\in A$, $P_{Z|X}(\cdot|x)$ is 
a probability mass function on $\{0,1\}$,
and where both $P_{Z|X}(0|x)$ and $P_{Z|X}(1|x)$
are measurable functions of $x$.
The test result is now the {\em random} 
binary value of 
the random variable $Z$ which, conditional on $X=x$, has
distribution $P_{Z|X}(\cdot|x)$, $x\in A$.

The two error probabilities associated with a test $P_{Z|X}$,
\ben
e_1\;=\;e_1(P_{Z|X})&=&\BBP(Z=0|X\sim Q)=\int_{A}P_{Z|X}(0|x)\,dQ(x),\\
e_2\;=\;e_2(P_{Z|X})&=&\BBP(Z=1|X\sim P)=\int_{A}P_{Z|X}(1|x)\,dP(x),
\een
determine the test's performance.
The best achievable performance among all
deterministic or randomised tests can be described as the
smallest possible value, $e_1^*(\epsilon)$, of the first 
error probability $e_1$ over all tests whose
second error probability, $e_2$, satisfies $e_2\leq \epsilon$,
$$e_1^*(\epsilon):=\inf \big\{ e_1(P_{Z|X})\;:\; 
P_{Z|X}\;\;\mbox{s.t.}\;\; e_2(P_{Z|X})\leq\epsilon\big\}.$$

Here, our interest is in the case of hypothesis
tests between two product measures $P^n$ and~$Q^n$, 
corresponding to a sequence $X_1^n=(X_1,X_2,\ldots,X_n)$ 
of independent and identically (i.i.d.) observations with 
values in $A$. 
We write $e_{1,n}$ and $e_{2,n}$ for the error probabilities
of a specific test, and
we denote the best achievable value of the first error 
probability by:
$$e_{1,n}^*(\epsilon):=\inf\big\{e_{1,n}(P_{Z|X_1^n})\;:\;
P_{Z|X_1^n}\;\;\mbox{s.t.}\;\; e_{2,n}(P_{Z|X_1^n})\leq\epsilon\big\}.$$

\subsection{Background}

When the maximum allowed value $\epsilon>0$ of the second
error probability is fixed,
the first-order asymptotic behaviour of $e_{1,n}^*(\epsilon)$
is described by what has been come to be known 
as {\em Stein's lemma}~\cite{chernoff:52,kullback-book,cover:book2}.
It states that smallest achievable first error probability
$e_{1,n}^*(\epsilon)$
decays exponentially with the sample size~$n$,
$$\log e_{1,n}^*(\epsilon)= -nD(P\|Q)+o(n),
\quad\mbox{as}\;n\to\infty,$$
where all logarithms are taken to base~$e$ ($\log=\log_e$ throughout),
and $D(P\|Q)$ denotes the relative entropy
between two probability measures $P$ and $Q$ on $(A,\clA)$:
$$D(P\|Q):=
\begin{cases}
	\int\frac{dP}{dQ}\log\big(\frac{dP}{dQ}\big)\,dQ,
	\quad&\mbox{if}\;P\ll Q,\\
	+\infty,\quad&\mbox{otherwise}.
\end{cases}$$

Stein's lemma was refined by Strassen~\cite{strassen:64b} who
claimed that, as $n\to\infty$,
\bqq
\log e_{1,n}^*(\epsilon) = -nD(P\|Q) -\sqrt{n}\sigma\Phi^{-1}(\epsilon) 
- \frac{1}{2}\log n  + O(1), 
\label{eq:strassen}
\eqq
where $\sigma^2$ is the variance of the log-likelihood
ratio $\log[\frac{dP}{dQ}(X_1)]$ with $X_1\sim P$, and $\Phi$ denotes the
standard normal distribution function.
Although some issues of rigour were raised in~\cite{kontoyiannis-verdu:14}
regarding Strassen's proof, an even stronger
version of~(\ref{eq:strassen}) was established 
by Polyanskiy-Poor-Verd\'{u} in~\cite{PPV:10},
where explicit, finite-$n$ bounds were obtained for the $O(1)$
term.
For the sake of completeness, we state and prove a version
of these bounds in Theorems~\ref{thm:Dconverse_pol}
and~\ref{thm:Dachieve_pol}
in Section~\ref{s:prelim}.

When the maximum allowed value $\epsilon>0$ of the second
error probability is not fixed, but is required to 
decay to zero exponentially fast, it turns out that,
at least for a certain range of exponential rates, 
it is possible also have first probability of error
decay to zero at an exponential rate
as $n\to\infty$. Specifically,
suppose that the second probability of error is required
to be no greater than $e^{-n\delta}$, for some
$\delta>0$, and let $E_{1,n}^*(\delta)$
denote the best achievable $e_{1,n}$ error 
probability:
$$E^*_{1,n}(\delta):=e_{1,n}^*(e^{-n\delta}).$$
In this case,
Hoeffding~\cite{hoeffding:65} showed that, 
for any $\delta\in(0,D(Q\|P))$,
$$\log E^*_{1,n}(\delta)=-nD(\delta)+o(n),
\quad\mbox{as}\;n\to\infty,$$
where the optimal error exponent $D(\delta)$ is
given by,
\bqq
D(\delta):=\inf_{P':D(P'\|P)\leq\delta}D(P'\|Q).
\label{eq:Ddelta}
\eqq

A refinement of this, analogous to Strassen's refinement
of Stein's lemma, was claimed by 
Csisz{\'a}r and Longo~\cite{csiszar:71},
who stated that, 
\bqq
\log E^*_{1,n}(\delta)=-nD(\delta)-\frac{1}{2(1-\alpha^*)}\log n+O(1),
\label{eq:CsL}
\eqq
as $n\to\infty$,
where $\alpha^*$ is an appropriate constant in $(0,1)$ that 
depends on $P,Q$ and $\delta$. The precise asymptotic
limit of the $O(1)$ term was determined
in~\cite{dobrushin:62} and~\cite{vazquez:18}.

All of the works mentioned so far,
including the one by Csisz{\'a}r and Longo~\cite{csiszar:71},
are based on classical
bounds and expansions related to normal approximation 
in the regime of the central limit theorem --
namely, versions of the Berry-Ess\'{e}en bound and 
the first-order Edgeworth expansion~\cite{fellerII:book}.
A slightly different approach was recently taken in~\cite{vazquez:18}.
There, the authors employ saddlepoint approximations~\cite{jensen:book}
to derive precise expansions for both error
probabilities, showing that, at best, we can have,
\begin{align}
\log e_{1,n}
&=
	-nD(\delta)-\frac{1}{2}\log n+O(1),
	\label{eq:albert1}\\
\mbox{and}\quad \log e_{2,n}
&=
	-n\delta-\frac{1}{2}\log n+O(1),
	\quad\mbox{as}\;n\to\infty.
	\label{eq:albert2}
\end{align}
Although at first sight the expansions~(\ref{eq:albert1}) 
and~(\ref{eq:albert2}) might seem somewhat
different from~(\ref{eq:CsL}),
it is actually easy to show that they are very closely related. 

\subsection{Finite-n asymptotic approximation}

In the same sense in which the finite-$n$ bounds of
Polyanskiy-Poor-Verd\'{u}~\cite{PPV:10}
strengthen 
the asymptotic expansion~(\ref{eq:strassen})
of Strassen,
it is possible to derive 
{\em finite-$n$} bounds that provide a 
corresponding stronger version of the asymptotic 
expansions~(\ref{eq:CsL}),~(\ref{eq:albert1}),~(\ref{eq:albert2})
of~\cite{csiszar:71,vazquez:18}.
In Theorems~\ref{thm:ach_new} and~\ref{thm:Dachieve_large}
of Section~\ref{s:main} we provide explicit constants
$C$ and $C'$ such that,
\bqq
C'\leq
\log E_{1}^*(n, \delta) -\Big[ -nD(\delta) 
-\frac{1}{2(1-\alpha^*)}\log n\Big]
	\leq C,
\label{eq:main}
\eqq
for all $n$ greater 
than some explicit $N_0$.
These are exactly analogous to the
bounds derived in~\cite{kontoyiannis-verdu:14}
and~\cite{gavalakis-K:21}
for lossless data compression without and with side
information, respectively.

Lower bounds in the same spirit 
as our Theorem~\ref{thm:Dachieve_large} were 
previously established for finite alphabets
by Altu{\u{g}} and Wagner~\cite{altug:11,altug:14} 
and for general spaces by Nakibo{\u{g}}lu~\cite{nakibouglu:19}.
A general upper bound similar to that in our
Theorems~\ref{thm:ach_new} was more recently
also proved by Nakibo{\u{g}}lu~\cite{nakibouglu:20}.

Despite their technical nature,
the approximations to $e_{1,n}^*$ provided by all these different
approaches are of as much practical relevance as they 
are of mathematical interest. In particular, they each
are useful in different regimes of the sample-size
and the error probability requirements.
For example, Stein's lemma and its refined form in~(\ref{eq:strassen})
are only relevant for `moderate' values of $\epsilon$. 
If $\epsilon$ is small, then the second term dominates
and the approximation is no longer valid or useful.

For example, consider the case of 
observations generated by either 
$P \sim \Bern(0.6)$ or $Q \sim \Bern(0.25)$ on $A=\{0,1\}$.
In the case when $\epsilon$ is relatively small,
we examine the four approximations to $e_{1,n}^*(\epsilon)$
suggested by the above results. In the `Stein regime',
\begin{align}
\mbox{Stein:}\qquad &\log e_{1,n}^*\approx-nD(P\|Q),
	\label{eq:stein}\\
\mbox{Strassen:}\qquad &\log e_{1,n}^*\approx-nD(P\|Q)
	-\sqrt{n}\sigma\Phi^{-1}(\epsilon)-\frac{1}{2}\log n,
	\label{eq:strassen2}
\end{align}
and in the `error exponents regime',
\begin{align}
\mbox{Hoeffding:}
	\qquad &\log e_{1,n}^*\approx -nD(\delta)
	\label{eq:hoeffding}\\
\mbox{Thms.~\ref{thm:ach_new}-\ref{thm:Dachieve_large}:}
	\qquad &\log e_{1,n}^*\approx -nD(\delta)
	-\frac{1}{2(1-\alpha^*)}\log n.
	\label{eq:newapprox}
\end{align}
Table~\ref{Table:n=50} shows 
representative results for different small
values of $\epsilon$,
when the sample size $n = 50$.
These results clearly demonstrate the utility 
of the approximation~(\ref{eq:newapprox})
as it provides by far the most accurate estimates 
of the error probability $e_{1,n}^*(\epsilon)$
for the problem parameters considered.

\begin{table}[ht!]
\centering
\begin{tabular}{|c||c|c|c|c|c|}
\hline
	\multicolumn{1}{|c||}{$\epsilon$} 
	& \multicolumn{1}{|c|}{$e_{1,n}^*(\epsilon)$}
	& \multicolumn{1}{c|}{Stein~(\ref{eq:stein})} 
	& \multicolumn{1}{c|}{Strassen~(\ref{eq:strassen2})}
	& \multicolumn{1}{c|}{Hoeffding~(\ref{eq:hoeffding})} 
	& \multicolumn{1}{c|}
		{Thms.~\ref{thm:ach_new}-\ref{thm:Dachieve_large}:~(\ref{eq:newapprox})}\\
	\hline
 0.00006 & {\bf 0.098} & $10^{-6}$ & 84.2  & 0.804 & {\bf 0.083} \\
 0.00018 & {\bf 0.055} & $10^{-6}$ & 18.8  & 0.639 & {\bf 0.056} \\
 0.00052 & {\bf 0.029} & $10^{-6}$ & 4.23  & 0.472 & {\bf 0.035} \\
 0.00137 & {\bf 0.014} & $10^{-6}$ & 0.096 & 0.324 & {\bf 0.019} \\ 
 0.00336 & {\bf 0.006} & $10^{-6}$ & 0.022 & 0.208 & {\bf 0.010} \\ 
 0.00762 & {\bf 0.003} & $10^{-6}$ & 0.050 & 0.126 & {\bf 0.005} \\ 
 0.01604 & {\bf 0.001} & $10^{-6}$ & 0.011 & 0.071 & {\bf 0.002} \\ 
\hline
\end{tabular}
\centering
\caption{Comparison between the true value of the optimal error
probability $e_{1,n}^*(\epsilon)$ and four different approximations.
Clearly the approximation suggested by Theorems~\ref{thm:ach_new}
and~\ref{thm:Dachieve_large} gives the best results in this regime.}
\label{Table:n=50}
\end{table}

We close this introduction with a brief discussion 
of some related work of a different nature.
An important special case of binary hypothesis testing
is the fully symmetric regime, where the natural 
question is: If both
error probabilities are required to be no greater than some $\epsilon>0$,
how big does the sample size $n$ need to be in order for a test
to exist with this property? It turns 
out~\cite{borovkov:98,bar-yossef:02,canonne:20},
that the smallest such sample size,
$n^*(P,Q,\epsilon)$,~is,
$$n^*(P,Q,\epsilon)=\Theta\Big(\frac{\log(1/\epsilon)}{H^2_2(P,Q)}\Big),$$
where $H_2(P,Q)$ denotes the 
Hellinger distance between $P$ and $Q$.
This suggests the rough approximation,
\bqq
\log e_{1,n}^*(\epsilon)\approx -nH_2^2(P,Q).
\label{eq:samplec}
\eqq
But taking $e_{1,n}=e_{2,n}$, or equivalently 
$D(\delta)=\delta$ in the error exponents regime,
leads to an error exponent $C(P,Q)$ known
as the Chernoff information, which,
at least for $P$ close to $Q$, is approximately
equal to $H_2^2(P,Q)$. Therefore,
in the symmetric regime, the 
estimates~(\ref{eq:hoeffding}) 
and~(\ref{eq:samplec})
are roughly equivalent.

Finally, we mention that extensive discussions 
of different aspects of binary hypothesis testing
can be found in Csisz\'{a}r and K\"{o}rner's classic
text,~\cite{csiszar:book2},
Han's book on information spectrum methods~\cite{han:book-en},
and Tan's monograph~\cite{tan:book}.

\section{Preliminaries}
\label{s:prelim}

Consider a binary hypothesis test between two probability measures
$P,Q$ on $(A,\clA)$, absolutely continuous with respect
to a $\sigma$-finite dominating measure $\lambda$.
Let $p,q$ denote the corresponding densities,
$p = dP/d\lambda$ and $q = dQ/d\lambda$.
Although our main object
of interest, the optimal error probability $e_{1}^*(\epsilon)$,
cannot be explicitly evaluated in general, the optimal
test that achieves $e_1^*(\epsilon)$ is well known and given 
by the classical Neyman-Pearson region~\cite{neyman:33}. 
The following one-shot converse and achievability 
bounds are simple variations of the 
Neyman-Pearson lemma.  In this form, they were 
proved in~\cite{PPV:10}; the are stated here without proof.

\begin{proposition}[One-shot converse]
\label{prop:osconverse}
For any two probability measures $P$ and $Q$ on the 
same measurable space $(A, \cal{A})$,
for any test $P_{Z|X}$, and for any $\gamma>0$,
we have:
$$
e_1\geq\frac{1}{\gamma}\left[ 1-e_2-
	\BBP\Big(\log \frac{dP}{dQ}(X) > \log \gamma \Big)\right],
	\quad X\sim P.
$$
\end{proposition}

The following simple achievability result 
is obtained via deterministic tests.

\begin{proposition}[One-shot achievability]
\label{prop:achieve}
Let $P,Q$ be two probability measures on the 
same measurable space $(A, \cal{A})$, and
let $\epsilon\in(0,1)$. For any $\gamma>0$ such that,
$$
\BBP\Big(\log\frac{dP}{dQ}(X)<\log\gamma \Big)
\leq\epsilon,\quad X\sim P.
$$
we have,
$$e_1^*(\epsilon)\leq \BBP\Big(\log\frac{dP}{dQ}(Y)\geq\log\gamma \Big),
\quad Y\sim Q.
$$
\end{proposition}

Next, we apply the above one-shot bounds 
to the hypothesis testing problem with i.i.d.\ data
as described in the Introduction,
to derive refined, finite-$n$ approximations 
to $e_{1,n}^*(\epsilon)$
in the Stein regime.
Theorems~\ref{thm:Dconverse_pol} and~\ref{thm:Dachieve_pol}
are essentially proved in~\cite{PPV:10}. For the sake
of completeness, we include their proofs in the Appendix.

\begin{theorem}[Stein regime converse]
\label{thm:Dconverse_pol}
Suppose $P\neq Q$ are two probability measures on 
the same measurable space $(A,\cal{A})$,
such that:
$$T = \int \Big| \log\frac{dP}{dQ} - D(P\|Q)\Big|^3 \,dP<\infty.$$
Let $\epsilon\in(0,1)$.
Then, for any $\Delta > 0$, we have,
$$
        \log e_{1,n}^*(\epsilon) \geq -nD(P\|Q)
	- \sqrt{n}\sigma\Phi^{-1}\Big(\epsilon+\frac{B+\Delta}{\sqrt{n}}\Big) 
	-\frac{1}{2}\log n
	+\log\Delta,
$$
for all $n\geq [(B+\Delta)/(1-\epsilon)]^2$, where,
$$\sigma^2
= 
	\int
	\Big[ \log\frac{dP}{dQ} - D(P\|Q)\Big]^2 \,dP
\quad\mbox{and}\quad
B
= 
	\frac{T}{2\sigma^3}.
$$
\end{theorem}

\begin{theorem}[Stein regime achievability]
\label{thm:Dachieve_pol}
Suppose $P\neq Q$ are two probability measures 
on the same measurable space $(A,\cal{A})$,
and let $\epsilon\in(0,1)$.
Then, 
\begin{equation}
        \log e_{1,n}^*(\epsilon) \leq -nD(P\|Q)
	- \sqrt{n}\sigma\Phi^{-1}\Big(\epsilon-\frac{B}{\sqrt{n}}\Big) 
	-\frac{1}{2}\log n+ \log\Big(\frac{1}{\sqrt{2\pi}\sigma} +2B\Big),
        \label{eq:nachieve}
\end{equation}
for all $n>(B/\epsilon)^2$, where $\sigma,$ $B$ and $T<\infty$
are as in Theorem~\ref{thm:Dconverse_pol}.
\end{theorem}

In the following section we will need some simple results
about the error exponent function $D(\delta)$.
These are stated in Lemmas~\ref{lemma:useful_props} 
and~\ref{lemma:der} and Proposition~\ref{prop:alpha},
and they are proved in the Appendix.

For any two probability measures $P,Q$ on $(A, \cal{A})$ with 
densities $p,q$ respectively, 
for all $\alpha \in (0,1)$ we define:
\begin{equation}
    \label{eq:z_alpha_def} Z(\alpha) = \int p^{\alpha}q^{1-\alpha}\, d\lambda.
\end{equation}

\begin{lemma}[Properties of $Z(\alpha)$]
    \label{lemma:useful_props}
    \begin{itemize}
        \item[$(i)$]  If $D(P\|Q)$ and $D(Q\|P)$ are both finite,
	then $Z(\alpha)$ has a continuous derivative at 
	each $\alpha \in (0,1)$, with:
            \begin{equation}
                \label{eq:first_der_z_alpha}
                \frac{dZ(\alpha)}{d\alpha} 
	= \int p^{\alpha}q^{1-\alpha} \log \frac{p}{q} \,d\lambda.
            \end{equation}
        \item[$(ii)$]  If, moreover, 
	$\int p |\log(p/q)|^3 \,d\lambda$ 
	and $\int q |\log(p/q)|^3 \,d\lambda$ are both finite, 
	then $Z(\alpha)$ is three times differentiable 
	in $\alpha \in (0,1)$, with:
        \begin{equation}
            \label{eq:second_der_z}
            \frac{d^2 Z(\alpha)}{d\alpha^2}  
	= \int p^{\alpha}q^{1-\alpha} \Big[\log\frac{p}{q}  \Big]^2 \, d\lambda,
    \end{equation}
    \begin{equation}
        \label{eq:third_der_z}
        \frac{d^3 Z(\alpha)}{d\alpha^3}  
	= \int p^{\alpha}q^{1-\alpha} \Big[\log\frac{p}{q}  \Big]^3 \, d\lambda.
    \end{equation}
    \end{itemize}
\end{lemma}

\begin{lemma}[Derivatives of the error exponents]
    \label{lemma:der}
	For each $\alpha\in(0,1)$, let $Q_{\alpha}$ denote the
	probability measure on $(A,\clA)$ with density,
    \begin{equation}
        \label{eq:def_q_alpha}
        q_{\alpha} := \frac{p^{\alpha}q^{1-\alpha}}{Z(\alpha)},
    \end{equation}
    where $Z(\alpha)$ is defined in~{\em (\ref{eq:z_alpha_def})}. 
	Under the assumptions of Lemma~\ref{lemma:useful_props}~$(ii)$, 
	both $D(Q_{\alpha}\|P)$ and $D(Q_{\alpha}\|Q)$ are twice 
	differentiable in $\alpha\in(0,1)$  with:
    \begin{align}
\frac{dD(Q_{\alpha}\|P)}{d\alpha} 
&= 
	-(1-\alpha)\VAR
	\Big(\log\frac{dP}{dQ} (X^{(\alpha)})\Big),
        \label{eq:first_der_P}\\
\frac{dD(Q_{\alpha}\|Q)}{d\alpha} 
&= \alpha \VAR
	\Big(\log\frac{dP}{dQ}(X^{(\alpha)})\Big),
        \label{eq:first_der_Q}\\
\frac{d^2D(Q_{\alpha}\|P)}{d\alpha^2} 
&= 
	\VAR\Big(\log\frac{dP}{dQ}(X^{(\alpha)})\Big) 
	- (1-\alpha) 
	\BBE \Bigg[ \Bigg(\log\frac{dP}{dQ}(X^{(\alpha)})
	- \BBE\Big( \log\frac{dP}{dQ}(X^{(\alpha)}) 
	\Big)\Bigg)^3 \Bigg],
    	\label{eq:second_der_P}\\
\frac{d^2D(Q_{\alpha}\|Q)}{d\alpha^2} 
&= 
	\VAR\Big(\log\frac{dP}{dQ}(X^{(\alpha)})\Big) 
	+\alpha \BBE \Bigg[ \Bigg(\log\frac{dP}{dQ}(X^{(\alpha)}) 
	- \BBE\Big( \log\frac{dP}{dQ}(X^{(\alpha)})
	\Big)\Bigg)^3 \Bigg],
    	\label{eq:second_der_Q}
\end{align}
where $X^{(\alpha)} \sim Q_{\alpha}$.
\end{lemma}

Recall the definition of the optimal error
exponent $D(\delta)$ in~(\ref{eq:Ddelta}).
The following representation of $D(\delta)$ is
well-known for finite spaces $A$; 
see, e.g.,~\cite{csiszar:71} or~\cite{blahut:74}. 

\begin{proposition}[Error exponent representation]
\label{prop:alpha}
Let $P,Q$ be two probability measures on $(A,\clA)$
satisfying 
the assumptions of Lemma~\ref{lemma:useful_props}~$(ii)$.
Then
for any $0<\delta<D(Q\|P)$, there exists
a unique $\alpha^*\in(0,1)$ such that,
$D(Q_\alpha^*\|P)=\delta$,
$$D(\delta):=\inf_{P':D(P'\|P)\leq\delta} D(P'\|Q)
=D(Q_{\alpha^*}\|Q),
$$
and,
$$D(\delta)=\Big(\frac{\alpha^*}{1-\alpha^*}\Big)\delta
-\Big(\frac{1}{1-\alpha^*}\Big)\log Z(\alpha^*),$$
where $Z(\alpha)$ and $Q_{\alpha}$ are defined 
in~{\em (\ref{eq:z_alpha_def})} and~{\em (\ref{eq:def_q_alpha})}, 
respectively.
\end{proposition}


Parts of our main arguments later will be based on the
classical Berry-Ess\'{e}en bound~\cite{korolev:10,petrov-book:95}.

\begin{lemma}[Berry-Ess\'{e}en bound] 
\label{lem:BE}
Suppose $Z_1, Z_2, \ldots, Z_n$ are i.i.d.\ 
random variables with mean $\mu = \mathbb{E}(Z_1)$,
variance $\sigma^2 = \VAR(Z_1)$,
and $\rho = \mathbb{E}(|Z_1- \mu|^3) < \infty$.  
Let $\Bar{Z}_n$ denote the empirical
average, $\Bar{Z}_n = \frac{1}{n}(Z_1+Z_2 + \ldots +Z_n)$.
We have:
$$        \Bigg| \mathbb{P}\Bigg(   \frac{\sqrt{n}(\Bar{Z}_n - \mu)}{\sigma} 
	\leq x \Bigg) - \Phi(x)\Bigg| \leq \frac{\rho}{2\sigma^3\sqrt{n}},
	\quad x\in\RL,\; n\geq 1.
$$
\end{lemma}

The following is a simple consequence of the
Berry-Ess\'{e}en bound. It is proved in the Appendix.

\begin{lemma}
    \label{lem:BE2}
    With $Z_1, Z_2, \ldots, Z_n$ as in Lemma~\ref{lem:BE},
    we have:
    \begin{equation*}
        \BBE\big[
    e^{-\sum_{i=1}^n Z_i}
    \IND_{\{\sum_{i=1}^n Z_i\geq x\}} 
    \big]
    \leq
    \Big(\frac{1}{\sqrt{2\pi}} 
    + \frac{\rho}{\sigma^2}\Big)\frac{1}{\sqrt{n}\sigma}e^{-x},
    \quad x\in\RL,\;n\geq 1.
    \end{equation*}
\end{lemma}

\section{Finite-sample bounds}
\label{s:main}

We are now ready to state the main results,
stated in equation~(\ref{eq:main}) in the Introduction. 
Recall the definition of the optimal error
exponent $D(\delta)$ in~(\ref{eq:Ddelta})
and its representation in terms of the
family of `tilted' distributions $Q_{\alpha}$,
$\alpha\in(0,1)$ in Proposition~\ref{prop:alpha}.

\begin{theorem}[Achievability]
\label{thm:ach_new}
Let $P,Q$ be two probability measures on the measurable space 
$(A, \cal{A})$ with densities $p,q$, respectively,
with respect to the $\sigma$-finite dominating measure $\lambda$. 
Suppose that $\int p |\log(p/q)|^3 d\,\lambda$ 
and $\int q |\log(p/q)|^3 \,d\lambda$ are both finite. 
Then, for any $0<\delta<D(Q\|P)$ we have,
$$\log E_{1}^*(n, \delta) \leq
	-nD(\delta) -\frac{1}{2(1-\alpha^*)}\log n
	+C,$$
for all $n\geq 1$,
where the constant $C=C(\delta,P,Q)$ is given by,
\bqq
C= \log \Big(\frac{1}{\sigma^*\alpha^*\sqrt{2\pi}}  
	+ \frac{\rho^*}{\sigma^{*3}}\Big)
	+ \frac{\alpha^*}{1-\alpha^*}
	\log 
	\Big(\frac{1}{\sigma^*(1-\alpha^*)\sqrt{2\pi}}  
	+ \frac{\rho^*}{\sigma^{*3}}\Big),
\label{eq:C}
\eqq
with $\alpha^*$ being the unique $\alpha\in(0,1)$
that achieves
$\delta=D(Q_{\alpha^*}\|P)$, and where,
\begin{align*}
\sigma^{*2} 
&= 
	\VAR\Big(\log\frac{dQ_{\alpha^*}}{dP}(X^{(\alpha^*)})\Big),\\
\rho^* 
&= 
	\mathbb{E}\left| \log\frac{dQ_{\alpha^*}}{dP}(X^{(\alpha^*)})
	- \mathbb{E}\Big(\log\frac{dQ_{\alpha^*}}{dP}(X^{(\alpha^*)})
	\Big) \right|^3,
\end{align*}
with $X^{(\alpha^*)} \sim Q_{\alpha}$.
\end{theorem}

\noindent
{\sc Proof. } 
By Proposition~\ref{prop:alpha}, there exists a 
unique $\alpha^* \in (0,1)$ such that $D(Q_{\alpha^*}\|P) = \delta$.
Let $\gamma_n>0$ be a constant that will be chosen later,
and consider the decision region,
$$B_n
:=\Big\{\log\frac{p^n}{q^n}<\log\gamma_n\Big\}.$$
In view of the definition of $q_{\alpha^*}$, 
and using independence, $B_n$ can also be expressed
as,
\begin{align}
B_n
&=
	\Big\{ 
	\frac{1}{1-\alpha^*}\log\frac{p^n}{q^n_{\alpha^*}}
	< \frac{n\log Z_{\alpha^*}}{1-\alpha^*} +\log \gamma_n\Big\}
	\nonumber\\
&=
	\Big\{x_1^n\in A^n\;:\; 
	\sum_{i=1}^n\log\frac{q_{\alpha^*}(x_i)}{p(x_i)} >-n\log Z(\alpha^*)
	- (1-\alpha^*)\log \gamma_n\Big\}. 
    	\label{eq:ach_set_P}
\end{align}
Equivalently, interchanging the roles of $P$ and $Q$, we get,
\begin{equation}
    B_n = \Big\{x_1^n\in A^n\;:\; 
	\sum_{i=1}^n\log\frac{q_{\alpha^*}(x_i)}{q(x_i)} 
	<-n\log Z(\alpha^*) +\alpha^*\log \gamma_n\Big\}. 
    \label{eq:ach_set_Q}
\end{equation}

Now define,
$$\tau_1^2 
= \VAR\Big(\log\frac{dQ_{\alpha^*}}{dP}(X^{(\alpha^*)}) \Big)
\quad\mbox{and}\quad
r_1= \mathbb{E}\Bigg| 
\log\frac{dQ_{\alpha^*}}{dP}(X^{(\alpha^*)}) - \mathbb{E}
\Bigg(\log\frac{dQ_{\alpha^*}}{dP}(X^{(\alpha^*)})\Bigg) 
\Bigg|^3.$$
First, we claim that, for the choice
of $\gamma_n$ given by,
$$    \log \gamma_n := n(D(\delta) - \delta) +\frac{1}{2(1-\alpha^*)}\log n
	-\frac{1}{1-\alpha^*}\log 
	\left[\Big(\frac{1}{\tau_1\sqrt{2\pi}}  
	+ \frac{r_1}{\tau^{3}_1}\Big)\right],
$$
we have,
\bqq
\BBP\Big(\log\frac{dP^n}{dQ^n}(X_1^n)<\log\gamma_n\Big)\leq e^{-n\delta},
\label{eq:acondition}
\eqq
where $X_1^n\sim P^n$.
To show that, let
$(X^{(\alpha^*)}_1,X^{(\alpha^*)}_2,\ldots,X^{(\alpha^*)}_n)\sim
Q_{\alpha^*}^n$,
and write,
\begin{align*}
    \BBP\Big(\log\frac{dP^n}{dQ^n}(X_1^n)<\log\gamma_n\Big) &= \int_{A^n}
    \IND_{B_n}\,dP^n\\
&= \int_{A^n} 
	\exp\Bigg\{\sum_{i=1}^n\log\frac{p(x_i)}{q_{\alpha^*}(x_i)}\Bigg\}
	\IND_{B_n}\,dQ_{\alpha^*}^n\\
&=
	\BBE\Bigg[
	\exp\Bigg\{-\sum_{i=1}^n\log
	\frac{q_{\alpha^*}(X^{(\alpha^*)}_i)}{p(X^{(\alpha^*)}_i)}\Bigg\}
	\times\\
&
	\qquad
	\times
	\IND \Bigg\{ \sum_{i=1}^n\log
	\frac{q_{\alpha^*}(X^{(\alpha^*)}_i)}{p(X^{(\alpha^*)}_i)} 
	>-n\log Z(\alpha^*) - (1-\alpha^*)\log \gamma_n\Bigg\}
	\Bigg],
\end{align*}
where we used the representation of $B_n$ in~(\ref{eq:ach_set_P}).
Note that the terms in the above sums are i.i.d.\ with 
variance and absolute third centred moment given by $\tau_1^2$ and $r_1$,
respectively. Applying Lemma~\ref{lem:BE2}, then, yields,
$$     \BBP\Big(\log\frac{dP^n}{dQ^n}(X_1^n)<\log\gamma_n\Big) 
	\leq \frac{Z(\alpha^*)^n \gamma_n^{1-\alpha^*}}
	{ \sqrt{n}}\Big(\frac{1}{\tau_1\sqrt{2\pi}}  
	+ \frac{r_1}{\tau^3_1}\Big).
$$
And using the definition of $\gamma_n$ and Proposition~\ref{prop:alpha},
we obtain,
$$
\BBP\Big(\log\frac{dP^n}{dQ^n}(X_1^n)<\log\gamma_n\Big) \leq \exp \Big(  
-n(1-\alpha^*)[D(\delta)-\delta] +n(1-\alpha^*)D(\delta) -n\alpha^* \delta)
\Big) = e^{-n\delta},
$$
which is exactly~(\ref{eq:acondition}).

Given that $\gamma_n$ satisfies~(\ref{eq:acondition}), 
we can now apply the one-shot achievability result of
Proposition~\ref{prop:achieve} with $P^n,Q^n$ in place
of $P,Q$, to obtain that,
$$E^*_1(n, \delta)
\leq 
	\BBP\Big(\log\frac{dP^n}{dQ^n}(Y_1^n)\geq\log\gamma_n\Big)\\
=
	\int \IND_{B^c_n}\,dQ^n,
$$
with $Y_1^n\sim Q^n$, 
and using the expression for $B_n$ in~(\ref{eq:ach_set_Q}),
\begin{align*}
E_{1}^*(n, \delta)
&\leq
	\int_{A^n} 
	\exp\Bigg\{\sum_{i=1}^n\log\frac{q(x_i)}{q_{\alpha^*}(x_i)}\Bigg\}
	\IND_{B^c_n}(x_1^n)\,dQ_{\alpha^*}^n (x_1^n)\\
&=
\BBE\Bigg[
	\exp\Bigg\{\!-\sum_{i=1}^n\log\frac{q_{\alpha^*}(X^{(\alpha^*)}_i)}
	{q(X^{(\alpha^*)}_i)}\Bigg\}
	\IND \Bigg\{ \sum_{i=1}^n\log\frac{q_{\alpha^*}(X^{(\alpha^*)}_i)}
	{q(X^{(\alpha^*)}_i)} 
	\geq-n\log Z_{\alpha^*} +\alpha^*\log \gamma_n\Bigg\}
	\Bigg].
\end{align*}
Note that the sums above consist of i.i.d.\ terms with
variance and absolute
third centred moment given, respectively, by,
$$\tau_2^2 
= \VAR\Big(\log\frac{q_{\alpha^*}(X^{(\alpha^*)})}{q(X^{(\alpha^*)})} \Big)
\quad\mbox{and}\quad
r_2= \mathbb{E}\Bigg| 
\log \frac{q_{\alpha^*}(X^{(\alpha^*)})}{q(X^{(\alpha^*)})} - \mathbb{E}
\Bigg(\log \frac{q_{\alpha^*}(X^{(\alpha^*)})}{q(X^{(\alpha^*)})} \Bigg) 
\Bigg|^3.$$
Then applying Lemma~\ref{lem:BE2}, we can further bound,
$$
E_{1}^*(n, \delta)
\leq
	\frac{Z(\alpha^*)^n }{ \gamma_n^{\alpha^*}\sqrt{n}}
	\Big(\frac{1}{\tau_2\sqrt{2\pi}}  
	+ \frac{r_2}{\tau^3_2}\Big),
$$
and using the definition of $\gamma_n$ and Proposition~\ref{prop:alpha},
\begin{align}
\log E_{1}^*(n, \delta)
&\leq
	n\log Z(\alpha^*)-\alpha^*\log \gamma_n -\frac{1}{2}\log n 
	+\log \Big(\frac{1}{\tau_2\sqrt{2\pi}}  
	+ \frac{r_2}{\tau^3_2}\Big)
	\nonumber\\
&= 
	-nD(\delta) -\frac{1}{2(1-\alpha^*)}\log n
	\nonumber\\
&\qquad 
	+ 
	\log \Big(\frac{1}{\tau_2\sqrt{2\pi}}  
	+ \frac{r_2}{\tau^3_2}\Big) 
	+\frac{\alpha^*}{1-\alpha^*}
	\log \Big(\frac{1}{\tau_1\sqrt{2\pi}}  
	+ \frac{r_1}{\tau^3_1}\Big).
	\label{eq:pretarget}
\end{align}

Finally, we show that the sum of the two constants in~(\ref{eq:pretarget})
is exactly equal to $C$ in~(\ref{eq:C}).
Using the definition of $Q_{\alpha^*}$,
direct computation gives,
$$
\tau_1^2 
= 
	\VAR\Big(
	\log\frac{q_{\alpha^*}(X^{(\alpha^*)})}{p(X^{(\alpha^*)})} \Big)
= 
	\VAR\Big( 
	-(1-\alpha^*)\log \frac{p(X^{(\alpha^*)})}{q(X^{(\alpha^*)})} 
	-\log Z_{\alpha^*}  
	\Big),
$$
so that,
\bqq
\tau_1^2 
= 
	(1-\alpha^*)^2
	\VAR\Big(
	\log \frac{p(X^{(\alpha^*)})}{q(X^{(\alpha^*)})} \Big)
= 
	(1-\alpha^*)^2\sigma^{*2},
\label{eq:var_to_var}
\eqq
and similarly, we obtain,
\bqq
\tau_2^2 =\alpha^{*2}\sigma^{*2}.
\label{eq:var_to_var2}
\eqq
For the third moments $r_1$ and $r_2$, analogous
computations give,
\begin{align}
r_1 
&= 
	\mathbb{E}
	\Bigg| \log \frac{q_{\alpha^*}(X^{(\alpha^*)})}{p(X^{(\alpha^*)})} 
	- \mathbb{E}\Bigg(\log 
	\frac{q_{\alpha^*}(X^{(\alpha^*)})}{p(X^{(\alpha^*)})} \Bigg) 
	\Bigg|^3 
	\nonumber\\
&= 
	(1-\alpha^*)^3\mathbb{E}
	\Bigg| \log \frac{p(X^{(\alpha^*)})}{q(X^{(\alpha^*)})} - \mathbb{E}
	\Bigg(\log \frac{p(X^{(\alpha^*)})}{q(X^{(\alpha^*)})} \Bigg) 
	\Bigg|^3
	\nonumber\\
&=
	(1-\alpha^*)^3\rho^{*},
	\label{eq:rho_to_rho}
\end{align}
and similarly,
\begin{equation}
    r_2 = \alpha^{*3}\rho^{*}.
    \label{eq:rho_to_rho2}
\end{equation}
Substituting~(\ref{eq:var_to_var}),~(\ref{eq:var_to_var2}),~(\ref{eq:rho_to_rho}),
and~(\ref{eq:rho_to_rho2}) into~(\ref{eq:pretarget}),
completes the proof.
\qed

\begin{theorem}[Converse]
\label{thm:Dachieve_large}
Under the same assumptions and in the notation of Theorem~\ref{thm:ach_new}, 
suppose that,
$$\sigma_0^2 := \inf_{\alpha \in (0,1)} 
\VAR\Big(\log\frac{p(X^{(\alpha)})}{q(X^{(\alpha)})} \Big) > 0,
$$
and,
$$\rho_0 :=  \sup_{\alpha \in (0,1)} \mathbb{E}
\Bigg| \log \frac{p(X^{(\alpha)})}{q(X^{(\alpha)})} 
- \mathbb{E}\Bigg(\log \frac{p(X^{(\alpha)})}{q(X^{(\alpha)})} 
\Bigg) \Bigg|^3 <\infty,$$
where $X^{(\alpha)}\sim Q_\alpha$ with 
$Q_\alpha$ as in Proposition~\ref{prop:alpha}.

Then, 
\bqq
\label{thmnew_converse}
\log E_1^*(n, \delta) \geq -nD(\delta) 
- \frac{1}{2(1-\alpha^*)}\log n  +C',
\eqq
where the constant $C'=C'(\delta,P,Q)$ is given by,
$$C'=
\frac{\log 2}{(1-\alpha^*)} - \frac{(\sigma^{*2} + 2\rho_0)(2-\alpha^*)}
{2(1-\alpha^*)} -\frac{|\sigma_0^2-\rho_0|}{2}
+m,
$$
where,
$$m =  -2\sqrt{2\pi}(1-\alpha^*)
(\rho_0/\sigma_0^3+1)\sqrt{\sigma^{*2} + \rho_0},$$
and the bound~{\em (\ref{thmnew_converse})} holds
for all,
$$n> \max\Big\{7(\rho_0/\sigma_0^3+1)^2,  
\frac{(\sigma^{*2}+2\rho_0-2m+2\log2)^2}{(1-\alpha^*)^2\sigma^{*4}}, 
	n_0\Big\},$$
where
$n_0$ is the smallest $n$ such that $\log n 
\leq (1-\alpha^*)\sigma^{*2}\sqrt{n}.$
\end{theorem}

\noindent
{\sc Proof. }
With $A^n,P^n,Q^n$ in place of $A,P,Q$, respectively, 
and with $X_1^n\sim P^n$, the result will eventually 
follow by an application of 
Proposition~\ref{prop:osconverse} for a particular
choice of $\gamma_n$:
\bqq
E_{1}^*(n, \delta)
\geq \frac{1}{\gamma_n} 
\left[ \BBP\Big(\log \frac{dP^n}{dQ^n} (X_1^n)
\leq \log \gamma_n\Big) - e^{-n\delta}  \right].
\label{eq:ineq_conv1}
\eqq
Here we choose $\gamma_n$ via,
$$\log \gamma_n = n(D(\delta_n)-\delta_n),$$
where
$\delta_n=D(Q_{\alpha_n}\|P)$, and $\alpha_n$ is given defined
via the relation,
\bqq
n(\alpha_n- \alpha^*)(1-\alpha^*)\sigma^2
=
	\frac{1}{2}\log n
	+ \frac{1}{2}(\sigma^{*2}+2\rho_0)-q + \log 2,
\label{eq:alphan}
\eqq
so that,
$$
    \alpha_n 
	= \alpha^* 
	+ \frac{1}{2(1-\alpha^*)\sigma^{*2}}\times\frac{\log n}{n} 
	+ \frac{(\sigma^{*2}+2\rho_0)/2-m 
	+ \log 2}{(1-\alpha^*)\sigma^{*2}}\times \frac{1}{n},
$$
with $m,\rho_0$ and $\sigma_0^2$ as in the statement of theorem.
The reason for this choice is that,
as we show next, this $\alpha_n$
leads to a $\gamma_n$ such that,
\begin{equation}
\label{eq:converse_ineq_aim}
   \BBP\Big(\log \frac{dP^n}{dQ^n}(X_1^n)
\leq \log \gamma_n\Big) \geq 2e^{-n\delta}. 
\end{equation}
Establishing~(\ref{eq:converse_ineq_aim}) will occupy
most of this proof.

Let,
$$B_n
:=\Big\{x_1^n\in A^n\;
:\;\log\frac{p^n(x_1^n)}{q^n(x_1^n)}\leq \log\gamma_n\Big\},$$
and note that, as in the earlier proof of Theorem~\ref{thm:ach_new},
$B_n$ can also be expressed as,
\begin{align}
B_n 
&=
	\Bigg\{x_1^n\in A^n\;:\; 
	\log\frac{p^n(x_1^n)}{q_{\alpha_n}(x_1^n)} 
	\leq n\log Z(\alpha_n) +(1-\alpha_n)\log \gamma_n\Bigg\}
	\nonumber\\
&=
	\Bigg\{x_1^n\in A^n\;:\; 
	\log\frac{q^n_{\alpha_n}(x_1^n)}{p^n(x_1^n)} \geq 
	n\BBE\Big( \log\frac{q_{\alpha_n}(X^{(\alpha_n)})}{p(X^{(\alpha_n)})}  
	\Big)\Bigg\}.
	\label{eq:altB_n}
\end{align}
Using the Berry-Ess\'{e}en bound in Lemma~\ref{lem:BE},
we can bound above
the probability of $B^c_n$ under $Q_{\alpha_n}^n$ as,
\begin{equation*}
Q_{\alpha_n}^n(B_n^c) 
=  \BBP\Bigg(\sum_{i = 1}^n \log
\frac{q_{\alpha_n}(X^{(\alpha_n)}_i)}{p(X^{(\alpha_n)}_i)} 
< n\BBE\Big( \log
\frac{q_{\alpha_n}(X^{(\alpha_n)})}{p(X^{(\alpha_n)})} \Big) \Bigg)
\leq \Phi(0) + \frac{r_n}{2\tau_n^3\sqrt{n}},
\end{equation*}
where $(X_1^{(\alpha_n)},\ldots,X_n^{(\alpha_n)})\sim Q_{\alpha_n}^n$,
$$
\tau^{2}_n 
= \VAR\Big(\log\frac{q_{\alpha_n}(X^{(\alpha_n)})}{p(X^{(\alpha_n)})}\Big),
$$ 
and,
$$r_n 
= \mathbb{E} \Bigg| \log \frac{q_{\alpha_n}(X^{(\alpha_n)})}{p(X^{(\alpha_n)})} 
- \mathbb{E} \Bigg(\log \frac{q_{\alpha_n}(X^{(\alpha_n)})}{p(X^{(\alpha_n)})} 
\Bigg) \Bigg|^3.$$
Using the exact same computations that led to
relations~(\ref{eq:var_to_var}) and~(\ref{eq:rho_to_rho}) 
in the previous proof, here we have 
that $\tau_n = (1-\alpha_n)\sigma_n$ and $r_n = (1-\alpha_n)^3\rho_n$ 
where $\sigma_n^2$ and $\rho_n$ are the variance and the absolute third 
centred moment of $\log[p(X^{(\alpha_n)})/q(X^{(\alpha_n)})]$. Therefore,
\begin{equation}
\label{eq:ub_converse}
     Q_{\alpha_n}^n(B_n^c) 
	\leq 
	\frac{1}{2} + \frac{\rho_0}{2\sigma_0^3\sqrt{n}}.
\end{equation}

Now, with $P$ in place of $Q$ and 
$Q_{\alpha_n}$ in place of $P$,
and in view of~(\ref{eq:altB_n}),
Proposition~\ref{prop:achieve} implies that
the probability in~(\ref{eq:converse_ineq_aim}) 
can be bounded below as,
$$
\BBP\Big(\log \frac{p^n(X_1^n)}{q^n(X_1^n)} 
\leq \log \gamma_n\Big) 
\geq
\tilde{e}_{1,n}\big(Q_{\alpha_n}^n(B_n^c)\big),
$$
where 
$\tilde{e}_{1,n}(\epsilon)$ is the same as
the minimal error probability $e^*_{1,n}(\epsilon)$,
but with $P$ in place of $Q$ and $Q_{\alpha_n}$ in place of $P$.
Moreover, by~(\ref{eq:ub_converse}), 
$$
\BBP\Big(\log \frac{p^n(X_1^n)}{q^n(X_1^n)} 
\leq \log \gamma_n\Big) 
\geq
\tilde{e}_{1,n}\Big(
	\frac{1}{2} + \frac{\rho_0}{2\sigma_0^3\sqrt{n}}
\Big),
$$
and we can use the Stein regime converse,
Theorem~\ref{thm:Dconverse_pol}, with $\Delta=1$,
to obtain that,
\begin{align*}
\log\BBP 
&
	\Big(\log \frac{dP^n}{dQ^n}(X_1^n) \leq \log \gamma_n\Big) \\
&\geq 
	-nD(Q_{\alpha_n}\|P) 
	- \sqrt{n}\tau_n\Phi^{-1}\Big(\frac{1}{2} 
	+ \frac{\rho_0}{2\sigma_0^3\sqrt{n}} 
	+ \frac{r_n}{2\tau_n^3\sqrt{n}} + \frac{1}{\sqrt{n}}\Big) 
	-\frac{1}{2}\log n\\
&\geq  
	-nD(Q_{\alpha_n}\|P) - \sqrt{n}\tau_n\Phi^{-1}\Big(\frac{1}{2} 
	+ \frac{\rho_0}{\sigma_0^3\sqrt{n}}+ \frac{1}{\sqrt{n}}\Big) 
	-\frac{1}{2}\log n,
\end{align*}
for all $n > 4 (\rho_0/\sigma_0^3 + 1)^2$. 
Using the fact that $\alpha_n\geq\alpha^*$ by definition,
and substituting a simple two-term Taylor expansions for
$\Phi^{-1}$, gives,
\begin{align}
\label{eq:lower_intermediate}
\nonumber 
\log \BBP\Big(\log \frac{dP^n}{dQ^n}(X_1^n)
\leq \log \gamma_n\Big) &\geq -nD(Q_{\alpha_n}\|P)
	\!-\! \sqrt{n}\tau_n\Phi^{-1}\Big(\frac{1}{2}\Big) 
	\!-\!\frac{1}{2}\log n - \frac{(\rho_0/\sigma_0^3+1)\tau_n}{\phi(\Phi^{-1}(\frac{1}{2}+\frac{\rho_0/\sigma_0^3+1}{\sqrt{n}}))}\\
 &\geq  -nD(Q_{\alpha_n}\|P)
	-\frac{1}{2}\log n - \frac{(1-\alpha^*)(\rho_0/\sigma_0^3+1)\sigma_n}{\phi(\Phi^{-1}(\frac{1}{2}+\frac{\rho_0/\sigma_0^3+1}{\sqrt{n}}))}.
 \end{align}

There are two more terms
in~(\ref{eq:lower_intermediate})
that will be bounded using a 
second order Taylor expansion. 
For $\sigma_n$, 
using~(\ref{eq:second_der_P}) from Lemma~\ref{lemma:der}
we have,
\begin{align*}
\sigma_n^2 
&= 
	\VAR\Big(\log \frac{dP}{dQ}(X^{(\alpha_n)})\Big)\\
&= 
	\VAR\Big(\log \frac{dP}{dQ}(X^{(\alpha_n)}) 
	+ (\alpha_n-\alpha^*)\
    	\BBE \Bigg[ \Bigg(\log\frac{dP}{dQ}(X^{(\psi)}) 
	- \BBE \Big( \log\frac{dP}{dQ}(X^{(\psi)})  
	\Big)\Bigg)^3 \Bigg], 
\end{align*}
for some $\psi \in [\alpha^*, \alpha_n]$. Since the conditions 
in~Lemma \ref{lemma:der} are fulfilled, the above derivatives 
are justified. Thus,
\bqq
\sigma_n^2 
\leq \VAR\Big(\log \frac{dP}{dQ}(X^{(\alpha^*)})\Big) 
+\BBE \Bigg| \log\frac{dP}{dQ}(X^{(\psi)}) 
- \BBE\Big( \log\frac{dP}{dQ}(X^{(\psi)})  \Big) \Bigg|^3 \\
\leq \sigma^{*2} + \rho_0.
\label{eq:expansion_var}
\eqq
And for $D(Q_{\alpha_n}\|P)$, 
using the first and second derivatives of $D(Q_{\alpha}\|P)$ 
in~(\ref{eq:first_der_P}) and~(\ref{eq:second_der_P})
of Lemma~\ref{lemma:der}, we have,
\bqq
D(Q_{\alpha_n}\|P) = D(Q_{\alpha^*}\|P) 
- (\alpha_n-\alpha^*)(1-\alpha^*)\sigma^{*2} + 
\frac{1}{2}(\alpha_n-\alpha^*)^2
S_n(\psi),
\label{eq:Taylor_P}
\eqq
where,
$$S_n(\psi)=
\VAR\Big(\log\frac{dP}{dQ}(X^{(\psi)}) \Big) 
- (1-\psi) 
\BBE\Bigg[ \Bigg(\log\frac{dP}{dQ}(X^{(\psi)})  - 
\BBE\Big( \log\frac{dP}{dQ}(X^{(\psi)}) \Big)\Bigg)^3 \Bigg],
$$
for some (possibly different from above) $\psi \in [\alpha^*, \alpha_n]$. 
Arguing as in~(\ref{eq:expansion_var}), $S(\psi)$ and subsequently
$D(Q_{\alpha_n}\|P)$ can be 
bounded above as,
\begin{equation}
\label{eq:expansion_P}
    D(Q_{\alpha_n}\|P) \leq D(Q_{\alpha^*}\|P) - (\alpha_n-\alpha^*)(1-\alpha^*)\sigma^{*2} + \frac{1}{2}(\alpha_n-\alpha^*)^2 (\sigma^{*2} + 2\rho_0).
\end{equation}

Substituting the bounds~(\ref{eq:expansion_var}) and~(\ref{eq:expansion_P}) 
in~(\ref{eq:lower_intermediate}) yields,
\begin{align*}
\log \BBP\Big(\log \frac{dP^n}{dQ^n}(X_1^n) \leq \log \gamma_n\Big) 
\geq &
	- nD(Q_{\alpha^*}\|P) + n(\alpha_n-\alpha^*)(1-\alpha^*)\sigma^{*2} 
	\\
& 
	- \frac{n}{2}(\alpha_n-\alpha^*)^2 (\sigma^{*2} + 2\rho_0)
      	-\frac{1}{2}\log n
	\\
&
	- \frac{(1-\alpha^*)(\rho_0/\sigma_0^3+1)\sqrt{\sigma^{*2} 
	+ \rho_0}}{\phi(\Phi^{-1}(\frac{1}{2}
	+\frac{\rho_0/\sigma_0^3+1}{\sqrt{n}}))}.
\end{align*}
If we further take 
$n \geq (\rho_0/\sigma_0^3+1)^2/(\Phi(\sqrt{2\log2}) - 1/2)^2$, 
and recalling the definition of $m$,
we can further bound the last term above to obtain,
\begin{align}
&
	\log \BBP\Big(\log \frac{p^n(X_1^n)}{q^n(X_1^n)} \leq \log \gamma_n\Big)
	\nonumber\\
&\geq 
- n\delta + n(\alpha_n-\alpha^*)(1-\alpha^*)\sigma^{*2} 
	- \frac{n}{2}(\alpha_n-\alpha^*)^2 (\sigma^{*2} + 2\rho_0)
      -\frac{1}{2}\log n +m.
    \label{eq:conv_final_P2}
\end{align}
It is actually at this point where the exact form of $\alpha_n$
is justified; recalling its definition,
if we take $n$ large enough such that the following three 
conditions are satisfied,
\begin{align*}
&\frac{\log n}{\sqrt{n}} \leq (1-\alpha^*)\sigma^{*2},\\
&n \geq \frac{[(\sigma^{*2}+2\rho_0)-2m+2\log2]^2}
	{(1-\alpha^*)^2\sigma^{*4}},\\
& n > \frac{1}{(1-\alpha^*)^2},
\end{align*}
then $\alpha_n < 1$ and 
$n(\alpha_n-\alpha^*)^2 \leq 1$.
Using this, the third term in the right-hand side
of~(\ref{eq:conv_final_P2}) can be bounded by 
$(\sigma^{*2}+2\rho_0)/2$, 
and 
substituting the expression for $\alpha_n$
in~(\ref{eq:alphan}) into~(\ref{eq:conv_final_P2}),
yields,
$$\log \BBP\Big(\log \frac{dP^n}{dQ^n}(X_1^n) \leq \log \gamma_n\Big)
\geq-n\delta+\log2,$$
which is 
exactly~(\ref{eq:converse_ineq_aim}),
as desired.

Finally, we can return to the 
result of Proposition~\ref{prop:osconverse} 
in~(\ref{eq:ineq_conv1}), which, 
after using
the bound~(\ref{eq:converse_ineq_aim}) and the
definition of $\gamma_n$ becomes,
\begin{equation}
    \label{eq:final_E1}
    \log E_1^*(n, \delta) \geq -nD(Q_{\alpha_n}\|Q) + nD(Q_{\alpha_n}\|P) - n\delta.
\end{equation}

For the first term,
a three-term Taylor expansion around $\alpha^*$,
where $D(Q_{\alpha^*}\|Q)=D(\delta)$, gives,
\begin{align}
nD(Q_{\alpha_n}\|Q) - nD(\delta) 
&\leq
	n(\alpha_n-\alpha^*)\alpha^*\sigma^{*2} 
	+ \frac{n}{2}(\alpha_n-\alpha^*)^2 (\sigma^{*2} + 2\rho_0)
	\nonumber\\
&\leq 
	n(\alpha_n-\alpha^*)\alpha^*\sigma^{*2} 
	+ \frac{1}{2}(\sigma^{*2} + 2\rho_0)
	\nonumber\\
&\leq 
	\frac{\alpha^*}{(1-\alpha^*)}\frac{\log n}{2} 
	+ \frac{\alpha^* (\sigma^{*2}+2\rho_0)/2-\alpha^*m
	+ \alpha^*\log 2}{(1-\alpha^*)} + \frac{1}{2}(\sigma^{*2} + 2\rho_0)
	\nonumber\\
&= 
	\frac{\alpha^*}{(1-\alpha^*)}\frac{\log n}{2} 
	+ \frac{\alpha^*(\log 2-m)}{(1-\alpha^*)}
	+ \frac{1}{2(1-\alpha^*)}(\sigma^{*2} + 2\rho_0),
    \label{eq:final_taylor_Q}
\end{align}
where we used again the expression for $\alpha_n$
from~(\ref{eq:alphan}), 
and the fact that $n(\alpha_n-\alpha^*)^2\leq 1$.
For the second term,
using the second order Taylor expansion from~(\ref{eq:Taylor_P}), 
we can similarly bound,
\begin{align}
nD(Q_{\alpha_n}\|P)
&\geq 
	nD(Q_{\alpha^*}\|P) - n(\alpha_n-\alpha^*)(1-\alpha^*)\sigma^{*2} 
	+ \frac{n(\alpha_n-\alpha^*)^2}{2} (\sigma_0^2 - \rho_0)
	\nonumber\\
&\geq 
	n\delta - n(\alpha_n-\alpha^*)(1-\alpha^*)\sigma^{*2} 
	- \frac{1}{2} |\sigma_0^2 - \rho_0|
	\nonumber\\
&\geq 
	n\delta - \frac{\log n}{2} - \frac{\sigma^{*2}+2\rho_0}{2} 
	+\log 2 -m - \frac{1}{2} |\sigma_0^2 - \rho_0|.
    \label{eq:final_P}
\end{align}
Substituting the bounds~(\ref{eq:final_taylor_Q}) and~(\ref{eq:final_P}) 
into~(\ref{eq:final_E1}) yields,
\begin{align*}
\log E_1^*(n, \delta) 
& \geq 
	-nD(\delta) - \frac{\alpha^*}{(1-\alpha^*)}\frac{\log n}{2} 
	- \frac{\alpha^*}{(1-\alpha^*)}(\log 2-m) 
	-\frac{(\sigma^{*2} + 2\rho_0)}{2(1-\alpha^*)}\\
& 
	\qquad
	-\frac{\log n}{2} - \frac{(\sigma^{*2}+2\rho_0)}{2} + \log 2-m 
	-\frac{|\sigma_0^2 - \rho_0|}{2}\\
&=
	-nD(\delta) - \frac{\log n}{2(1-\alpha^*)}
	+ \frac{(1-2\alpha^*)(\log 2-m)}{(1-\alpha^*)} - \frac{(\sigma^{*2} 
	+ 2\rho_0)(2-\alpha^*)}{2(1-\alpha^*)} 
	-\frac{|\sigma_0^2 - \rho_0|}{2},
\end{align*}
and using the definition of $m$ in the statement of the theorem
yields exactly the bound claimed in~(\ref{thmnew_converse}) 
and completes the proof.
\qed

\vspace{0.1in}

\appendix

\centerline{\LARGE\bf Appendix}

\medskip

\noindent
{\sc Proof of Theorem~\ref{thm:Dconverse_pol}. }
Our starting point is the one-shot converse in 
Proposition~\ref{prop:osconverse}.
With $A^n,P^n,Q^n$ in place of $A,P,Q$, respectively,
and with $X_1^n\sim P^n$, 
we have that, for any $\gamma_n>0$:
\bqq
e_{1,n}^*(\epsilon)
\geq \frac{1}{\gamma_n} 
\left[1-\epsilon - \BBP\Big(\log \frac{dP^n}{dQ^n} (X_1^n)
>\log \gamma_n\Big) \right],
\label{eq:nconverse3}
\eqq
with $X_1^n\sim P^n$.
To further bound $e_{1,n}^*(\epsilon)$,
we examine the probability in the right-hand side
of~(\ref{eq:nconverse3}). First, 
\begin{align}
\BBP\Big( \log \frac{dP^n}{dQ^n}(X_1^n) >\log \gamma_n \Big) 
&=
    	\BBP\Big(\sum_{i=1}^n\Big[\log \frac{dP}{dQ}(X_i)-D\Big] 
	>\log \gamma_n -nD\Big)
	\nonumber\\
& = 
	1- \BBP\Big(
	\frac{1}{\sqrt{n}\sigma}
	\sum_{i=1}^n\Big[\log \frac{dP}{dQ}(X_i) -  D\Big]
	\leq\frac{1}{\sqrt{n}\sigma}[\log \gamma_n - nD] \Big)
	\nonumber\\
& =
	1- F_n\Big(\frac{\log\gamma_n -nD}{\sqrt{n}\sigma}\Big),
	\label{eq:nconverse4}
\end{align}
where we write 
$F_n$ for 
the distribution function of
$\frac{1}{\sqrt{n}\sigma}\big(\sum_{i=1}^n[\log\frac{dP}{dQ}(X_i)-D]\big)$,
and $D$ for $D(P\|Q)$.
By the Berry-Ess\'{e}en bound in Lemma~\ref{lem:BE},
with $Z_i=\log[\frac{dP}{dQ}(X_i)]$,
we have,
\bqq
\Phi 
\Big(\frac{\log\gamma_n-nD}{\sqrt{n}\sigma}\Big)
- \frac{T}{2\sigma^3\sqrt{n}}
\leq
F_n
\Big(\frac{\log\gamma_n-nD}{\sqrt{n}\sigma}\Big)
\leq 
\Phi 
\Big(\frac{\log\gamma_n-nD}{\sqrt{n}\sigma}\Big)
+ \frac{T}{2\sigma^3\sqrt{n}},
\label{eq:nconverse5}
\eqq
and combining~(\ref{eq:nconverse3}) with~(\ref{eq:nconverse4})
and with the left-hand inequality in~(\ref{eq:nconverse5}),
\begin{align*}
\log e_{1,n}^*(\epsilon)
\geq -\log \gamma_n
+\log\Bigg(-\epsilon-\frac{B}{\sqrt{n}}
+\Phi\Big(\frac{\log\gamma_n-nD}{\sqrt{n}\sigma}\Big)
 \Bigg).
\end{align*}

Finally we will choose an appropriate value of $\gamma_n$.
Taking $\gamma_n= \exp\{nD +\sqrt{n}\sigma\Phi^{-1}(\psi_n)\}$ 
for some $\psi_n \in (0,1/2]$, yields,
\begin{align*}
\log e_{1,n}^*(\epsilon) 
\geq 
-nD -\sqrt{n}\sigma\Phi^{-1}(\psi_n)
+\log \Big(-\epsilon - \frac{B}{\sqrt{n}}  +\psi_n \Big),
\end{align*}
and choosing
$\psi_n = \epsilon + \frac{B+\Delta}{\sqrt{n}}$ 
for some $\Delta >0$,
\begin{align*}
\log e^*_{1,n} \geq  
-nD- \sqrt{n}\sigma
\Phi^{-1}\Big(\epsilon + \frac{B+\Delta}{\sqrt{n}}\Big) 
-\frac{1}{2} \log n +\log\Delta.
\end{align*}
This holds as long as $\psi_n$ can be chosen appropriately,
that is, as long as we have $\psi_n\in(0,1)$
so that the inverse function
$\Phi^{-1}$ is well defined. We obviously
always have $\psi_n>0$, and for $\psi_n<1$ we
need $n >[(B+\Delta)/(1-\epsilon)]^2$. This 
completes the proof of~(\ref{thm:Dconverse_pol}). 
\qed

\noindent
{\sc Proof of Theorem~\ref{thm:Dachieve_pol}. }
Our starting point now is the one-shot achievability result
in Proposition~\ref{prop:achieve}. With $A^n,P^n,Q^n$
in place of $A,P,Q$, respectively, 
let $X_1^n\sim P^n$, $Y_1^n\sim Q^n$, and
let $\gamma_n>0$
be such that,
\bqq
\BBP\Big(\log\frac{dP^n}{dQ^n}(X_1^n)<\gamma_n\Big)\leq\epsilon.
\label{eq:acondition2}
\eqq
Then Proposition~\ref{prop:achieve} states
that,
$$e_{1,n}^*(\epsilon)\leq \BBP\Big(\log\frac{dP^n}{dQ^n}(Y_1^n)
\geq\gamma_n\Big).$$ 
Writing $B_n\subset A^n$ for the decision region,
$$B_n
:=\Big\{\log\frac{dP^n}{dQ^n}\geq\log\gamma_n\Big\}
=\Big\{x_1^n\in A^n\;:\;\sum_{i=1}^n
\log\frac{dP}{dQ}(x_i)\geq\log\gamma_n\Big\},$$
we have,
\begin{align*}
e_{1,n}^*(\epsilon)
&\leq
	\int\IND_{B_n}\,dQ^n\\
&=
	\int
	\exp\Bigg\{-\sum_{i=1}^n\log\frac{dP}{dQ}(x_i)\Bigg\}
	\IND_{B_n}(x_1^n)\, dP^n(x_1^n)\\
&=
	\BBE\Bigg[
	\exp\Bigg\{-\sum_{i=1}^n\log\frac{dP}{dQ}(X_i)\Bigg\}
	\IND_{B_n}(X_1^n)
	\Bigg],
\end{align*}
and by Lemma~\ref{lem:BE2},
\bqq
e_{1,n}^*(\epsilon)
\leq
	\frac{1}{\gamma_n\sqrt{n}}\Big(\frac{1}{\sigma\sqrt{2\pi}}  
	+ 2B\Big).
\label{eq:nachieve3}
\eqq
Now, we choose a specific $\gamma_n$ and verify that~(\ref{eq:acondition2})
is satisfied. As in the proof of Theorem~\ref{thm:Dconverse_pol}, 
write $D=D(P\|Q)$ and
let $\gamma_n=nD+\sqrt{n}\sigma\Phi^{-1}(\psi_n)$,
but this time with $\psi_n=\epsilon-B/\sqrt{n}$.
From~(\ref{eq:nconverse5}) in the proof of Theorem~\ref{thm:Dconverse_pol}
we have that,
$$
\BBP\Big(\log\frac{dP^n}{dQ^n}(X_1^n)<\gamma_n\Big)
\leq F_n\Big(\frac{\log\gamma_n-nD}{\sqrt{n}\sigma}\Big)
\leq \Phi\Big(\frac{\log\gamma_n-nD}{\sqrt{n}\sigma}\Big)+\frac{B}{\sqrt{n}}
=\epsilon,
$$
so that~(\ref{eq:acondition2}) is satisfied, as long as $\psi_n\in(0,1)$,
i.e., as long as $n>(B/\epsilon)^2$.
And substituting the value of $\gamma_n$ in~(\ref{eq:nachieve3}) 
and taking logarithms,
yields exactly~(\ref{thm:Dachieve_pol}).
\qed

\noindent
{\sc Proof of Lemma~\ref{lemma:useful_props}.}
Formally, all three results of the lemma are
straightforward computations;
the only thing we need to justify is the interchange 
of the order of differentiation and integration.

Write
$h=\log\frac{p}{q}$. 
Since $D(P\|Q)$ and $D(Q\|P)$ are both
finite,
we have that $\int |h|\,dP$ and $\int |h|\,dQ$ are both
finite. Moreover, we have the simple inequality
$0\leq p^\alpha q^{1-\alpha}\leq p+q$ for all $\alpha\in(0,1)$.
Therefore, the expression for the first derivative in
the lemma is finite
for all $\alpha\in(0,1)$.
Also, the two measures $P,Q$ are mutually
absolutely continuous, so we can restrict the integral
in the definition of $Z(\alpha)$ to
$S:=\{p>0\}\cap\{q>0\}$. 

Now let $\alpha\in(0,1)$. For all
$\epsilon$ with $0<|\epsilon|<\min\{\alpha,1-\alpha\}$,
we have,
$$\frac{Z(\alpha+\epsilon)-Z(\alpha)}{\epsilon}
=\int_S\frac{1}{\epsilon}\big[p^{\alpha+\epsilon}q^{1-\alpha-\epsilon}
-p^{\alpha}q^{1-\alpha}
\big]\,d\lambda
=\int_S qe^{\alpha h}\Big[\frac{e^{\epsilon h}-1}{\epsilon}\Big]
\,d\lambda.
$$
By a first-order Taylor expansion,
the integrand above is equal to,
$$p^{\alpha+\xi}q^{1-\alpha-\xi}h,$$
for some $\xi=\xi(\epsilon)$ with $|\xi|\leq \epsilon$,
which 
is dominated by the $\lambda$-integrable function
$(p+q)|h|$, so the desired differentiation
under the integral is justified by dominated convergence.

To see that the derivative is continuous in $\alpha$, 
take $\alpha\in(0,1)$ and
$0<|\epsilon|<\min\{\alpha,1-\alpha\}$. Then,
$$Z'(\alpha+\epsilon)
=\int_S p^{\alpha+\epsilon}q^{1-\alpha-\epsilon}
h\,d\lambda,$$
where the integrand converges pointwise to 
$p^\alpha q^{1-\alpha} h$ as $\epsilon\to 0$,
and it is absolutely dominated by 
$(p+q)|h|$, which is $\lambda$-integrable.
Therefore, by dominated convergence,
$Z'(\alpha+\epsilon)\to Z'(\alpha)$ as
$\epsilon\to 0$, completing the proof of~$(i)$.

For~$(ii)$ we similarly note that since 
$\int ph^2\,d\lambda$ and 
$\int qh^2\,d\lambda$ 
are both finite, and since for all $\alpha$
we have $0\leq p^\alpha q^{1-\alpha}\leq p+q$,
the expression for the second derivative in the lemma is finite.
Then for $\alpha\in(0,1)$ and
$0<|\epsilon|<\min\{\alpha,1-\alpha\}$,
$$\frac{Z'(\alpha+\epsilon)-Z'(\alpha)}{\epsilon}
=\int_S\frac{1}{\epsilon}\big[p^{\alpha+\epsilon}q^{1-\alpha-\epsilon}
-p^{\alpha}q^{1-\alpha}
\big]h\,d\lambda
=\int_S qe^{\alpha h}\Big[\frac{e^{\epsilon h}-1}{\epsilon}\Big]h
\,d\lambda,
$$
where the integrand can be expressed as
$p^{\alpha+\xi'}q^{1-\alpha-\xi'}h^2,$
for some $\xi'=\xi'(\epsilon)$ with $|\xi'|\leq \epsilon$.
Since this is dominated by the $\lambda$-integrable function
$(p+q)h^2$, the exchange of differentiation and integration
is again justified by dominated convergence, 
justifying differentiating twice under the integral
sign.

The proof for the third derivative is perfectly analogous
to the two cases above.
\qed

\noindent
{\sc Proof of Lemma~\ref{lemma:der}. }
Starting with the definition of the density \ $q_{\alpha}$,
we have,
\begin{align*}
    D(Q_{\alpha}\|P) 
&= \int q_{\alpha} \log\Bigg( \frac{p^{\alpha}q^{1-\alpha}}{Z(\alpha)p}\Bigg) 
	\, d\lambda \\
    &=-\log Z(\alpha) -(1-\alpha) \int q_{\alpha} \log \frac{p}{q} 
	\, d\lambda\\
    &= -\log Z(\alpha) -\frac{1-\alpha}{Z(\alpha)} 
	\int p^{\alpha}q^{1-\alpha} \log \frac{p}{q} \, d\lambda.
\end{align*}
By Lemma~\ref{lemma:useful_props},
the last expression
above is differentiable 
in $\alpha\in (0,1)$, and can be expressed as,
$$D(Q_{\alpha}\|P) = -\log Z(\alpha) -\frac{(1-\alpha)Z'(\alpha)}{Z(\alpha)}.$$
Therefore, 
\begin{align*}
    \frac{d}{d\alpha} D(Q_{\alpha}\|P) &= -(1-\alpha) \frac{Z(\alpha)Z''(\alpha)-Z'(\alpha)^2}{Z(\alpha)^2}\\
    & = -(1-\alpha) \Big[\int q_{\alpha} h^2 \, d\lambda - \Big(\int q_{\alpha} h \, d\lambda \Big)^2 \Big],
\end{align*}
where $h = \log \frac{p}{q}$.  
This proves~(\ref{eq:first_der_P}). Again by Lemma~\ref{lemma:useful_props}
we have that this is differentiable, so that,
\begin{align*}
    \frac{d^2}{d\alpha^2} D(Q_{\alpha}\|P) &= \Big[\frac{Z''(\alpha)}{Z(\alpha)} - \Big( \frac{Z'(\alpha)}{Z'(\alpha)} \Big)  \Big] - (1-\alpha)\Big[ \frac{Z'''(\alpha)}{Z(\alpha)} -3\frac{Z''(\alpha)Z'(\alpha)}{Z(\alpha)^2} + 2\Big(\frac{Z'(\alpha)}{Z(\alpha)}\Big)^3\Big] \\
     &= \Big[\int q_{\alpha} h^2 \, d\lambda -
     \Big(\int q_{\alpha} h \, d\lambda \Big)^2 \Big]-\\ &-(1-\alpha)\Big[ \int q_{\alpha} h^3 \, d\lambda - 3 \int q_{\alpha} h^2 \, d\lambda \int q_{\alpha} h \, d\lambda + 2\Big( \int q_{\alpha} h \, d\lambda\Big)^3\Big],
\end{align*}
where the second term in the expression is the centered third moment of $\log\frac{p}{q}$ under $Q_{\alpha}$.
This proves~(\ref{eq:second_der_P}).
The proofs of~(\ref{eq:first_der_Q}) and~(\ref{eq:second_der_Q})
are identical.
\qed

\noindent
{\sc Proof of Proposition~\ref{prop:alpha}.}
Choose and fix a $0<\delta<D(Q\|P)$.
Let $g(\alpha)=D(Q_\alpha\|P)$, $\alpha\in[0,1]$.
Since $g(0)=D(Q|P)$ and $g(1)=0$,
to prove the existence of an $\alpha^*\in(0,1)$ 
with $g(\alpha^*)=\delta$ it suffices 
to show that 
$g(\alpha)$ is continuous on $[0,1]$.

Let $h=\log(p/q)$.
First we observe that $Z(\alpha)$ is bounded above
and below away from zero:
As noted earlier, we have
$p^\alpha q^{1-\alpha}\leq p+q$ 
for all $\alpha\in[0,1]$, and hence $Z(\alpha)\leq 2$,
$\alpha\in[0,1]$. Also, since 
$D(P\|Q)$ and $D(Q\|P)$ are both
finite, $P,Q$ are mutually
absolutely continuous, and
we must have that $m:=Q(p\geq q)>0$.
Then, for $\alpha\in(0,1)$:
$$Z(\alpha)
=\int p^\alpha q^{1-\alpha}\,d\lambda
\geq \int_{\{p\geq q\}} p^\alpha q^{1-\alpha}\,d\lambda
\geq \int_{\{p\geq q\}} q\,d\lambda = m>0.$$
Recalling Lemma~\ref{lemma:der}, we have,
\begin{align}
g(\alpha)
&=
	\int q_\alpha\log\frac{q_\alpha}{p}\,d\lambda
	\nonumber\\
&=
	\int q_\alpha\Big[\log\frac{1}{Z(\alpha)}
	+(1-\alpha)\log\frac{q}{p}\Big]\,d\lambda
	\nonumber\\
&=
	\log\frac{1}{Z(\alpha)}
	-
	(1-\alpha)\frac{Z'(\alpha)}{Z(\alpha)}.
	\label{eq:expandg}
\end{align}
so $g(\alpha)$ is continuous for $\alpha\in(0,1)$.
For all $\alpha\in[0,1]$ the last integrand
above is absolutely bounded by
$|p^\alpha q^{1-\alpha} h|\leq p|h|+q|h|,$
where the first absolute moments
$\int p|h|\,d\lambda$ and $\int q|h|\,d\lambda$
are both finite.
Hence, by dominated convergence,
the integral converges to $\int qh\,d\lambda=-D(Q\|P)$
as $\alpha\downarrow 0$. Moreover, dominated convergence
implies that $Z(\alpha)\to 1$ 
and hence $g(\alpha)\to g(0)=D(Q\|P)$ as
$\alpha\downarrow 0$.
Similarly, we have that 
$\int p^\alpha q^{1-\alpha}h\,d\lambda \to D(P\|Q)<\infty$
as $\alpha\uparrow 1$, which, combined with the fact that,
again by dominated convergence, we have
$Z(\alpha)\to 1$ as $\alpha\uparrow 1$, shows that
$g(\alpha)\to g(1)=0$ as $\alpha\uparrow 1$.

Therefore, $g$ is continuous on all of $[0,1]$, 
and the existence of an $\alpha^*\in(0,1)$
such that $D(Q_{\alpha^*}\|P)=\delta$ follows.
If moreover $\int p (\log\frac{p}{q})^2 \, d\lambda$ and $\int q (\log\frac{p}{q})^2 \, d\lambda$ are both finite, then, by Lemma~\ref{lemma:der} the first derivative of $g(\alpha)$ exists and is strictly negative for all $\alpha\in(0,1)$, hence proving the uniqueness of $\alpha^* \in (0,1)$.

Next, we prove that $D(\delta) = D(Q_{\alpha^*}\|Q)$. Of course, 
by definition,  $D(\delta) \leq D(Q_{\alpha^*}\|Q)$, so we only
need to prove that,
$$
D(W\|Q)\geq D(Q_{\alpha^*}\|Q),\;
\mbox{for any}\;W\;\mbox{with}\; D(W\|P)\leq \delta.$$
Indeed, let $W$ be a probability measure on $(A, \cal{A})$ such 
that $D(W\|P)\leq\delta<\infty$. Therefore, $W \ll P$ and 
\begin{align*}
(1-\alpha^*)[D(W\|Q)-D(Q_{\alpha^*}\|Q)]
&=
    D(W\|Q_{\alpha^*})
    -\alpha^* D(W\|P)
    +\alpha^* D(Q_{\alpha^*}\|P)\\
& \geq D(W\|Q_{\alpha^*}) \geq 0,
\end{align*}
where we used $D(W\|P)\leq \delta$ and $\delta = D(Q_{\alpha^*}\|P)$.\qed

\noindent {\sc Proof of Lemma~\ref{lem:BE2}. }
Let $F_n$ and $G_n$, respectively,
denote the distribution functions of $S_n$
and $[S_n-n\mu]/(\sigma\sqrt{n})$ where $S_n = \sum_{i=1}^n Z_i$.
Using integration by parts and the definition
of $G_n$,
\begin{align*}
\BBE\big[ e^{-S_n} \IND_{\{S_n\geq x\}}  \big]
&=
    \int_x^\infty e^{-s}\,dF_n(s)\\
&=
    -e^{-x}F_n(x)+\int_x^\infty e^{-s}F_n(s)\,ds\\
&=
    -e^{-x}G_n\Big(\frac{x-n\mu}{\sigma\sqrt{n}}\Big)+
    \int_x^\infty e^{-s}
    G_n\Big(\frac{s-n\mu}{\sigma\sqrt{n}}\Big)
    \,ds,
\end{align*}
and applying the Berry-Ess\'{e}en bound in Lemma~\ref{lem:BE},
\begin{align*}
\BBE\big[ e^{-S_n} \IND_{\{S_n\geq x\}}  \big]
&\leq
    -e^{-x}\Big[\Phi\Big(\frac{x-n\mu}{\sigma\sqrt{n}}\Big)
    -\frac{\rho}{2\sigma^3\sqrt{n}}\Big]
    + \int_x^\infty e^{-s}
    \Big[\Phi\Big(\frac{s-n\mu}{\sigma\sqrt{n}}\Big)
    +\frac{\rho}{2\sigma^3\sqrt{n}}\Big]\,ds\\
&=
    \int_x^\infty
    \Phi\Big(\frac{s-n\mu}{\sigma\sqrt{n}}\Big)
    e^{-s}\,ds
    -\Big[\Phi\Big(\frac{x-n\mu}{\sigma\sqrt{n}}\Big)
    -\frac{\rho}{\sigma^3\sqrt{n}}\Big]e^{-x}.
\end{align*}
Finally integrating by parts again,
\begin{align*}
\BBE\big[ e^{-S_n} \IND_{\{S_n\geq x\}}  \big]
&\leq
    \frac{1}{\sigma\sqrt{n}}\int_x^\infty e^{-s}
    \phi\Big(\frac{s-n\mu}{\sqrt{n}}\Big)\,ds
    +\frac{\rho}{\sigma^3\sqrt{n}}e^{-x}\\
&\leq
    \frac{1}{\sigma\sqrt{n}}e^{-x}\frac{1}{\sqrt{2\pi}}
    +\frac{\rho}{\sigma^3\sqrt{n}}e^{-x}\\
&=
    \Big(\frac{1}{\sqrt{2\pi}}+\frac{\rho}{\sigma^2}\Big)
    \frac{1}{\sigma\sqrt{n}}e^{-x},
\end{align*}
where we used the fact that the standard normal density $\phi$
is bounded above by $1/\sqrt{2\pi}$.
\qed

\bibliography{ik}
\bibliographystyle{plain}

\end{document}